\documentclass[aps,twocolumn,prd,preprintnumbers, 10pt]{revtex4-1}
\pdfoutput=1
\usepackage[hidelinks]{hyperref}
\usepackage{amssymb}
\usepackage{amsfonts}
\usepackage{graphicx}
\usepackage{epstopdf}
\usepackage{dcolumn}
\usepackage{amsmath}
\usepackage{latexsym,bm}
\usepackage{amsthm}
\usepackage{slashed}
\usepackage{float}
\usepackage{color}
\usepackage{url}
\usepackage{longtable}
\usepackage{rotating}
\usepackage[normalem]{ulem}
\usepackage{faktor}
\usepackage{tikz-cd}
\usepackage{tensor}
\usepackage{array}

\newcommand{\be}{\begin{equation}}
\newcommand{\ee}{\end{equation}}
\newcommand{\ba}{\begin{aligned}}
\newcommand{\ea}{\end{aligned}}

\newcommand{\Neu}{\text{Neu}}
\newcommand{\Dir}{\text{Dir}}

\definecolor{DarkGreen}{rgb}{0.1, 0.7, 0.3}

\renewcommand{\Vec}{\mathsf{Vec}}
\newcommand{\Rep}{\mathsf{Rep}}

\renewcommand{\dim}{\text{dim}}
\newcommand{\Bsym}{\mathfrak{B}^{\text{sym}}}
\newcommand{\Bphys}{\mathfrak{B}^{\text{phys}}}

\def\fT{\mathfrak{T}}
\def\fZ{\mathfrak{Z}}
\def\cA{\mathcal{A}}

\newcommand{\lid}{\mathbf{1}}
\newcommand{\lE}{\mathsf{E}}
\newcommand{\lP}{\mathsf{P}}
\newcommand{\lS}{\mathsf{S}}
\newcommand{\lA}{\mathsf{A}}
\newcommand{\Ising}{\mathsf{Ising}}
\newcommand{\TY}{\mathsf{TY}}
\newcommand{\Q}{\boldsymbol{Q}}

\newcommand{\phys}{\text{phys}}

\newcommand{\bit}{\begin{itemize}}
\newcommand{\eit}{\end{itemize}}
\newcommand{\ben}{\begin{enumerate}}
\newcommand{\een}{\end{enumerate}}

\newcommand{\id}{\text{id}}

\newcommand{\ot}{\otimes}
\newcommand{\half}{\frac{1}{2}}

\newcommand{\Z}{{\mathbb Z}}

\newcommand{\bC}{{\mathbb C}}

\newcommand{\cO}{\mathcal{O}}

\newcommand{\cS}{\mathcal{S}}

\newcommand{\cZ}{\mathcal{Z}}

\def\half{{\frac{1}{2}}}

\def\unit{{1\kern-.65ex {\rm l}}}
\def\1{{1\kern-.65ex {\rm l}}}

\def\bbZ{{\mathbb{Z}}}

\newcommand{\beq}{\begin{equation}}
\newcommand{\eeq}{\end{equation}}

\usepackage{verbatim}
\usepackage{tikz}
\usetikzlibrary{arrows,snakes,shapes.arrows,decorations.markings}
     \tikzset{>=triangle 90}
     \tikzstyle{gr}=[draw,circle,green!50!black,fill=green!50!black,scale=.6]
     \tikzstyle{Bl}=[draw,circle,blue,scale=.7]
     \tikzstyle{R}=[draw,circle,fill=red,scale=.7]
     \tikzstyle{bl}=[draw,circle,fill=black,scale=.2]
     \tikzstyle{bbc}=[draw,circle,fill=black,scale=.75]
     \tikzstyle{bbcs}=[draw,circle,fill=black,scale=.5]
     \tikzstyle{rc}=[circle,fill=red,scale=.6]
     \tikzstyle{wc}=[draw,circle,scale=.75]
\usepackage{array} % ??
\usepackage{multirow} % multiple row elements in a table
\usepackage{colortbl} % colored rows and columns in tables
\usepackage{stfloats}  % ??
\usepackage{cellspace}
\usepackage{xcolor}   % \color{...}, colored text
\newcommand{\xdasharrow}[2][->]{
\tikz[baseline=-\the\dimexpr\fontdimen22\textfont2\relax]{
\node[anchor=south,font=\scriptsize, inner ysep=1.5pt,outer xsep=2.2pt](x){#2};
\draw[shorten <=3.4pt,shorten >=3.4pt,dashed,#1](x.south west)--(x.south east);
}
}%\newcolumntype{C}[1]{%
\def\^{\wedge}

\def\Z{\mathbb{Z}}

\def\cA{{\mathcal A}}

\def\cO{{\mathcal O}}

\def\cS{{\mathcal S}}

\newcount\hour \newcount\minute
\hour=\time \divide \hour by 60
\minute=\time
\def\now{%
\ifnum \hour<13
  \ifnum \hour=0 \advance \hour by 12 \number\hour:\else \number\hour:\fi%
     \ifnum \minute<10 0\fi%
     \number\minute%
\ A.M.%
\else \advance \hour by -12 \number\hour:%
  \ifnum \minute<10 0\fi%
  \number\minute%
  \ P.M.%
\fi%
}

\usepackage{tikz}
\usetikzlibrary{arrows}
\usetikzlibrary{shapes.geometric,calc,arrows, positioning,shapes.misc,decorations.markings}
\tikzset{
  big arrow/.style={
    decoration={markings,mark=at position 1 with {\arrow[scale=2,#1]{>}}},
    postaction={decorate},
    shorten >=0.4pt},
  big arrow/.default=black}
\usetikzlibrary{external}
\usetikzlibrary{positioning}
\usetikzlibrary{calc}
\tikzset{gauge-node/.style={shape=circle, draw, minimum width=.6cm}}

\pgfdeclarelayer{edgelayer}
\pgfdeclarelayer{nodelayer}
\pgfsetlayers{edgelayer,nodelayer,main} 
\tikzstyle{none}=[inner sep=0pt] 

\tikzstyle{NodeCross}=[draw, shape=circle, cross out, inner sep=0pt, minimum size=6pt,line width=0.25mm]
\tikzstyle{Circle}=[draw, shape=circle, black, inner sep=0pt, minimum size=6pt]
\tikzstyle{rtriangle}=[fill=black, regular polygon, regular polygon sides=3, rotate=90, inner sep=0pt, minimum size=8pt]
\tikzstyle{ltriangle}=[fill=black, regular polygon, regular polygon sides=3, rotate=270, inner sep=0pt, minimum size=8pt]
\tikzstyle{rtriangleblue}=[fill={rgb,255: red,17; green,160; blue,255}, regular polygon, regular polygon sides=3, rotate=90, inner sep=0pt, minimum size=8pt]
\tikzstyle{ltriangleblue}=[fill={rgb,255: red,17; green,160; blue,255}, regular polygon, regular polygon sides=3, rotate=270, inner sep=0pt, minimum size=8pt]
\tikzstyle{rtrianglegreen}=[fill={rgb,255: red,69; green,255; blue,28}, regular polygon, regular polygon sides=3, rotate=90, inner sep=0pt, minimum size=8pt]
\tikzstyle{ltrianglegreen}=[fill={rgb,255: red,69; green,255; blue,28}, regular polygon, regular polygon sides=3, rotate=270, inner sep=0pt, minimum size=8pt]
\tikzstyle{Uprtriangle}=[fill=black, regular polygon, regular polygon sides=3, rotate=0, inner sep=0pt, minimum size=8pt]
\tikzstyle{Downltriangle}=[fill=black, regular polygon, regular polygon sides=3, rotate=180, inner sep=0pt, minimum size=8pt]
\tikzstyle{rtriangleAmber}=[fill={rgb,255: red, 191; green, 144; blue, 63}, regular polygon, regular polygon sides=3, rotate=90, inner sep=0pt, minimum size=8pt]
\tikzstyle{UprtriangleViolett}=[fill={rgb,255: red,255; green,0; blue,0}, regular polygon, regular polygon sides=3, rotate=0, inner sep=0pt, minimum size=8pt]

\tikzstyle{Downltriangle}=[fill=black, regular polygon, regular polygon sides=3, rotate=180, inner sep=0pt, minimum size=8pt]
\tikzstyle{UpRighttriangle}=[fill=black, regular polygon, regular polygon sides=3, rotate=45, inner sep=0pt, minimum size=8pt]
\tikzstyle{UpLefttriangle}=[fill=black, regular polygon, regular polygon sides=3, rotate=315, inner sep=0pt, minimum size=8pt]
\tikzstyle{DownRighttriangle}=[fill=black, regular polygon, regular polygon sides=3, rotate=135, inner sep=0pt, minimum size=8pt]
\tikzstyle{DownLighttriangle}=[fill=black, regular polygon, regular polygon sides=3, rotate=225, inner sep=0pt, minimum size=8pt]

\tikzstyle{Star}=[draw, shape=star, fill=black, star points=8, inner sep=0pt, minimum size=8pt]

\tikzstyle{DashedLine}=[-, densely dashed, line width=0.25mm]
\tikzstyle{DashedLineBrown}=[-, densely dashed, line width=0.25mm, draw={rgb,255: red,155; green,103; blue,51}]
\tikzstyle{DashedLineFall}=[-, densely dashed, line width=0.25mm, draw={rgb,255: red,195; green,0; blue,0}]
\tikzstyle{DashedLineViolett}=[-, densely dashed, line width=0.25mm, draw={rgb,255: red,139; green,41; blue,148}]
\tikzstyle{DottedLine}=[-, dotted, line width=0.25mm]
\tikzstyle{BlueLine}=[-, fill=none, draw={rgb,255: red,17; green,160; blue,255}, line width=0.25mm]
\tikzstyle{GreenLine}=[-, fill=none, draw={rgb,255: red,69; green,255; blue,28}, line width=0.25mm]
\tikzstyle{RedLine}=[-, draw={rgb,255: red,191; green,0; blue,0}, fill=none, line width=0.25mm]
\tikzstyle{DashedLineRed}=[-, densely dashed, fill=none, draw={rgb,255: red,191; green,0; blue,0}, line width=0.25mm]
\tikzstyle{ThickLine}=[-, line width=0.25mm]
\tikzstyle{ViolettLine}=[-, draw={rgb,255: red,132; green,60; blue,191}, fill=none, line width=0.25mm]
\tikzstyle{ViolettDashedLine}=[-, densely dashed, draw={rgb,255: red,132; green,60; blue,191}, fill=none, line width=0.25mm]
\tikzstyle{AmberLine}=[-, draw={rgb,255: red,191; green,144; blue,63}, fill=none, line width=0.25mm]
\tikzstyle{DashedRedThick}=[-, densely dashed, fill=none, draw={rgb,255: red,191; green,0; blue,0}, line width=0.40mm]
\tikzstyle{DashedBlueThick}=[-, densely dashed, fill=none, black, line width=0.40mm]

\makeatother

\begin{document}

\title{Categorical Landau Paradigm for Gapped Phases}

\author{Lakshya Bhardwaj}
\author{Lea E.\ Bottini}
\author{Daniel Pajer}
\author{Sakura Sch\"afer-Nameki}

\affiliation{Mathematical Institute, University
of Oxford, Woodstock Road, Oxford, OX2 6GG, United Kingdom}

%\date{\today}

\begin{abstract} % ≤ 600 characters!!! Wow!
\noindent 
We propose a unified framework to classify gapped infra-red (IR) phases with categorical symmetries, leading to a generalized, categorical Landau paradigm. This is applicable in any dimension and gives a succinct, comprehensive, and computationally powerful approach to classifying gapped symmetric phases.
The key tool is the symmetry topological field theory (SymTFT), which is a one dimension higher TFT with two boundaries, which we choose both to be topological.  We illustrate the general idea for (1+1)d gapped phases with categorical symmetries and suggest higher-dimensional extensions.
\end{abstract}

\keywords{Generalized Symmetries, Topological Phases, Confinement}

%%%%%%%%%%%%%%%%%%%%%%%%%%%%%%%%%

\maketitle

%\tableofcontents

\noindent{\bf Introduction.}
Uncovering the properties of quantum field theories (QFTs) and quantum matter systems in the far IR, \footnote{In this paper, we use terminology borrowed from the study of relativistic QFTs, where infrared corresponds to macroscopic distances and ultraviolet corresponds to microscopic distances.}
including their vacuum structure, hinges profoundly on the study of symmetries. Starting with the standard Landau theory, which applies to symmetries that form groups, the study of phases has had a long and successful history. 
The simplest setting is that of an operator, charged under a symmetry, acquiring a non-trivial vacuum expectation value (vev). The vev acts as an order parameter for the spontaneous symmetry breaking, and for gapped phases, this constrains the number of gapped vacua. 
However, as is well-known, not all phases can be described in this way. One generalization is to consider symmetries that go beyond groups, so-called categorical symmetries, which  have been actively researched in both condensed matter and high-energy physics, most notably in 
 $d>2$ starting with \cite{Heidenreich:2021xpr,
Kaidi:2021xfk, Choi:2021kmx, Roumpedakis:2022aik,Bhardwaj:2022yxj, Choi:2022zal} (for reviews see
\cite{Schafer-Nameki:2023jdn, Shao:2023gho}).

In this paper, we propose a general framework to study the impact of a categorical symmetry $\cS$ in the ultra-violet (UV) on the IR physics, focusing on gapped, i.e.\ topological, phases. 
Phases of matter are fundamental in condensed matter, but also in high-energy physics, as IR descriptions of QFTs. 
We determine the generalized charges under $\cS$ \cite{Bhardwaj:2023ayw} of order parameters for the gapped phases, leading to a categorical Landau paradigm \footnote{The Landau paradigm involves also the study of phase transitions between gapped phases. We expect similar properties of the generalized Landau paradigm being proposed here, with $\cS$-symmetric phase transitions arising by tuning the order parameters for the $\cS$-symmetric gapped phases discussed here} \footnote{Ideas like this have been around for a long time. This terminology was pointed out to us has appeared before in http://categorified.net/OSU.pdf and 
 https://scgp.stonybrook.edu/video/video.php?id=4960.}.
A classification of gapped $\cS$-symmetric phases will be achieved using the SymTFT. 
We illustrate this in $(1+1)$d, which has numerous applications in condensed matter physics, and $(3+1)$d with applications in confinement in QFTs. 
Other applications are to (2+1)d topological order, and spin liquids, as well as lattice models with categorical symmetries.
The $d>2$ case will be developed further in upcoming papers \cite{BBPSd}.

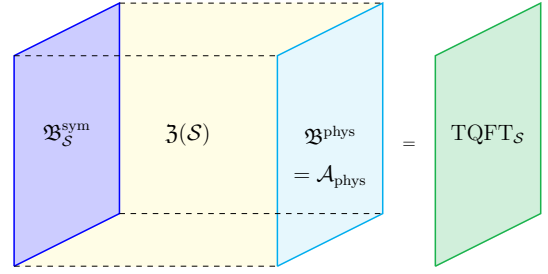
\begin{figure}[h!]
\centering
$
\scalebox{0.7}{
\begin{tikzpicture}
\draw [yellow, fill= yellow, opacity =0.1]
(0,0) -- (0,4) -- (2,5) -- (7,5) -- (7,1) -- (5,0)--(0,0);
\begin{scope}[shift={(0,0)}]
\draw [white, thick, fill=white,opacity=1]
(0,0) -- (0,4) -- (2,5) -- (2,1) -- (0,0);
\end{scope}
\begin{scope}[shift={(5,0)}]
\draw [white, thick, fill=white,opacity=1]
(0,0) -- (0,4) -- (2,5) -- (2,1) -- (0,0);
\end{scope}
\begin{scope}[shift={(5,0)}]
\draw [cyan, thick, fill=cyan,opacity=0.1] 
(0,0) -- (0, 4) -- (2, 5) -- (2,1) -- (0,0);
\draw [cyan, thick]
(0,0) -- (0, 4) -- (2, 5) -- (2,1) -- (0,0);
\node at (1,2.5) {\large ${\Bphys }$};
\node at (1,1.7) {\large ${ = \cA_\phys}$};
\end{scope}
\begin{scope}[shift={(0,0)}]
\draw [blue, thick, fill=blue,opacity=0.2]
(0,0) -- (0, 4) -- (2, 5) -- (2,1) -- (0,0);
\draw [blue, thick]
(0,0) -- (0, 4) -- (2, 5) -- (2,1) -- (0,0);
\node at (1,2.5) {\large $\Bsym_{\cS}$};
\end{scope}
\node at (3.3, 2.5) {\text{\large$\mathfrak{Z}({\mathcal{S}})$}};
\draw[dashed] (0,0) -- (5,0);
\draw[dashed] (0,4) -- (5,4);
\draw[dashed] (2,5) -- (7,5);
\draw[dashed] (2,1) -- (7,1);
\begin{scope}[shift={(8,0)}]
\node at (-0.5,2.3) {$=$} ;
\draw [DarkGreen, thick, fill=DarkGreen,opacity=0.2] 
(0,0) -- (0, 4) -- (2, 5) -- (2,1) -- (0,0);
\draw [DarkGreen, thick]
(0,0) -- (0, 4) -- (2, 5) -- (2,1) -- (0,0);
\node at (1,2.5) {\large TQFT${}_{\cS}$};    
\end{scope}
\end{tikzpicture}
}
$
\caption{SymTFT for Gapped Phases: the SymTFT $\mathfrak{Z} (\cS)$ is a $(d+1)$ dimensional topological field theory with two boundaries, which for gapped phases are  {\bf both topological}: $\Bsym_\cS = \cA_\cS$ and $\Bsym_{\phys}= \cA_\phys$.  \label{fig:SymTFT}}
\end{figure}

Many works have studied gapped theories with categorical symmetries in $d=2$ \footnote{For some recent works see \cite{Huang:2021zvu,Komargodski:2020mxz,Inamura:2021szw,Inamura:2021wuo,Chang:2018iay,Thorngren:2019iar, Chang:2018iay,Lin:2022dhv, Chatterjee:2022tyg} and a comprehensive comparison to the existing literature will be provided in \cite{Bhardwaj:2023idu}.}. Here, let us emphasize which aspects make the current proposal general and conceptually appealing, as well as computationally accessible: the approach is general but also provides refined data on gapped phases, beyond the number of vacua -- the action of the symmetry on the vacua, the order parameters (OPs), which have distinct features for non-invertible symmetries, and relative Euler terms -- and, perhaps most importantly, it is generalizable to higher dimensions. We show that spontaneous breaking of non-invertible symmetries can lead to physically distinguishable vacua: a physical phenomenon impossible for standard group symmetries. 

\noindent{\bf Generalized Symmetries.}
The main recent theoretical discovery has been the existence of generalized, not necessarily group-like, symmetries in $d>2$ QFTs. This generalization hinges on the identification of symmetries with topological defects \cite{Gaiotto:2014kfa}. The general structure is that of a fusion higher-category, which in first approximation means there are topological defects $D_p$ of dimensions $p = 0, \cdots, d-1$, which fuse according to 
\be
D_p^a \otimes D_p^b = \oplus_c N_{ab}^c D_p^{c}\,, \qquad N_{ab}^c \in \Z_+\,.
\ee
Finite, group-like symmetries arise when $N_{ab}^c = \delta_{c, ab}$ with $a,b,c\in G$ and $p=d-1$ \footnote{Allowing for more general $p$ but keeping the condition $N_{ab}^c = \delta_{c, ab}$ leads to higher-form and higher-group symmetries, whose study kickstarted the field of generalized symmetries.}.

\noindent{\bf The SymTFT.}
When studying categorical symmetries it is particularly useful to invoke the SymTFT  \cite{Ji:2019jhk, Gaiotto:2020iye, Apruzzi:2021nmk, Freed:2022qnc}, which separates the physical theory 
from its symmetries, allowing us to infer theory-independent aspects of the latter. It also provides a unified framework to study symmetries that are related by (generalized) gauging, and it encodes all the generalized charges  \cite{Bhardwaj:2023wzd, Bhardwaj:2023ayw} (i.e.\ local and extended operators that are charged under the categorical symmetry). In particular, two theories related by gauging global symmetries have the same SymTFT.

The SymTFT is a $(d+1)$-dimensional TQFT $\fZ(\cS)$ for a $d$-dimensional theory $\fT$ with a categorical symmetry $\cS$. It can be constructed by gauging the symmetry $\cS$ in $(d+1)$ dimensions. It has two boundaries: a topological boundary $\Bsym_\cS$, which encodes the symmetry $\cS$, and a not necessarily topological boundary $\Bphys_\fT$, which is theory specific and encodes its dynamics.
Compactifying the interval direction recovers $\fT$. 
So far the SymTFT has been used to study symmetry-related questions of physical (not necessarily topological) QFTs. 
 
\noindent{\bf Symmetric Gapped Phases from the SymTFT.}
As we are interested in \textit{gapped} phases, or  TQFTs, with $\cS$ symmetry, 
we will take the physical boundary to also be a topological boundary condition (b.c.).
This SymTFT set-up is shown in figure \ref{fig:SymTFT}. 
Classifying $\cS$-symmetric $d$-dimensional gapped phases, $d=D+1$ \footnote{Throughout the paper, we refer to TQFTs in $d$ space-time dimensions as being $d$-dimensional to emphasize the Euclidean nature of the space-time involved. On the other hand, we refer to gapped phases in $d=D+1$ space-time dimensions as being $(D+1)$-dimensional to emphasize the separate role played by the time direction.}, 
by utilizing the SymTFT perspective requires the following steps:
\smallskip

\noindent
 {\bf \emph{(1) SymTFT and Drinfeld Center:}} Given a symmetry category  $\cS$, we construct the associated $(d+1)$d SymTFT $\fZ(\cS)$ as in \cite{Bhardwaj:2023ayw}, which amounts to gauging $\cS$ in $(d+1)$d. The SymTFT has topological defects, which form the so-called Drinfeld center  $\cZ(\cS)$ of $\cS$, and play a key physical role as generalized charges. 

\noindent
 {\bf \emph{(2) Lagrangian Algebras:}} We then classify all the irreducible topological b.c.s of $\fZ(\cS)$, which are captured by Lagrangian algebras in $\cZ(\cS)$. These identify the topological defects of $\cZ(\cS)$ which can end (i.e.\ have Dirichlet b.c.s) on a given boundary.

\noindent
  {\bf \emph{(3)  Symmetry b.c.:}} 
    To classify $\cS$-symmetric gapped phases, we fix the symmetry boundary to be
    \be
\Bsym_\cS = \cA_{\cS}\,,
    \ee
    where $\cA_{\cS}$ is a Lagrangian algebra that realizes the symmetry $\cS$ on the boundary (generated by the topological defects with Neumann b.c.s). 

\noindent
 {\bf \emph{(4) Physical b.c.:}} 
The key difference with the standard SymTFT is the choice of physical boundary $\Bphys$, which we also take to be  topological, and is thus specified by a Lagrangian algebra $\cA_\phys$
\be
\Bphys=\cA_\phys\,,
\ee
which determines the topological defects that can end on $\Bphys$. 
Upon interval compactification of this SymTFT, we obtain a $d$-dimensional  TQFT.  By varying $\cA_\phys$, while keeping $\cA_\cS$ fixed, we move between different irreducible $\cS$-symmetric phases. 

\noindent
 {\bf \emph{(5) Generalized Charges as Order Parameters:}} 
For an arbitrary $\cS$-symmetric QFT $\fT$, the charges of (extended) operators under $\cS$ are captured by topological defects of the SymTFT that can end on its physical boundary $\Bphys_\fT$ \cite{Bhardwaj:2023ayw}. 
In our context, this implies that $\cA_{\phys}$ determines the charges of the operators in the $d$-dimensional theory under $\cS$:  $\cA_{\phys}$ captures the order parameters for the $\cS$-symmetric gapped phase.

The order parameters (OPs) for non-invertible symmetries will typically be a mixture of untwisted (conventional OP) and twisted-sector (string OP) operators, which combine to form irreducible multiplets under $\cS$ \footnote{By `twisted-sector' operator, we mean an operator at the end of a topological defect of $\cS$.}. 
This provides a generalized, categorical Landau paradigm describing gapped phases for an arbitrary categorical symmetry $\cS$.

\noindent{\bf Classification of (1+1)d Gapped Phases. }
We now specialize to unitary fusion categories $\cS$ in (1+1)d to provide a concrete implementation of the proposal. 
In this case we can extend the above program with additional refined properties for general $\cS$: 

\noindent
 {\bf \emph{(6)$^{(1+1)d}$ Vacua:}} 
 The number of vacua is easily determined by the number of the lines $\Q_i$ that can  end on both boundaries, i.e.\ that appear in both  $\cA_\cS$ and $\cA_\phys$: 
\begin{equation}\label{eq:basic_setup}
\begin{tikzpicture}
\begin{scope}[shift={(0,0)}]
\draw [thick] (0,-1) -- (0,1) ;
\draw [thick] (2,-1) -- (2,1) ;
\draw [thick] (2,0.6) -- (0,0.6) ;
\draw [thick] (0, -0.6) -- (2, -0.6) ;
\node[above] at (1,0.6) {$\Q_{1}$};
\node[above ] at (1,-0.6) {$\Q_{n}$};
\node[above] at (0,1) {$\cA_{\cS}$}; 
\node[rotate=90] at (1, 0.27) {$\cdots$};
\node[above] at (2,1) {$\cA_\phys$}; 
\draw [black,fill=black] (2,0.6) ellipse (0.05 and 0.05);
\draw [black,fill=black] (2,-0.6) ellipse (0.05 and 0.05);
\draw [black,fill=black] (0,0.6) ellipse (0.05 and 0.05);
\draw [black,fill=black] (0,-0.6) ellipse (0.05 and 0.05);
\end{scope}
\end{tikzpicture} 
\end{equation}

\noindent
 {\bf \emph{(7)$^{(1+1)d}$ Action of the Symmetry $\cS$:}}
The action of the symmetry $\cS$ on the (1+1)d gapped phase under discussion is specified by line operators $D_1^{(a)}$ of the associated 2d TQFT for each object $a\in\cS$, which represent the fusion category $\cS$ on the phases.  
The lines $D_1^{(a)}$ are determined as combinations of line operators of the 2d TQFT that act on the IR local operators realizing the order parameters according to their charges under $\cS$.

\noindent
 {\bf \emph{(8)$^{(1+1)d}$ SSB of Non-Invertibles and Euler Terms:}}
A notable phenomenon arises for (1+1)d gapped phases with categorical symmetries: the different vacua may be physically distinguishable as they can carry different Euler terms. Such terms are encoded in the properties of interfaces (which are line defects in 2d) between different vacua. Two vacua carrying different Euler terms are necessarily related by a line $D_1^{(a)}$ implementing a non-invertible symmetry $a\in\cS$ on the gapped phase, which is thus spontaneously broken. 
We conclude that: 
Spontaneous breaking of non-invertible symmetries can lead to physically distinguishable vacua. This is one hallmark of categorical symmetries.

\smallskip

\noindent{\bf Examples in ${(1+1)d}$.} 
This framework is applicable to {any (unitary) fusion category symmetry $\cS$}, which we now exemplify for invertibles and some non-invertible ones. 

\paragraph{\bf $\cS$ finite group.} For a non-anomalous group symmetry $G$, with 
$\cS = \Vec_{G}$, it is well known that (1+1)d gapped phases are a mixture of spontaneously symmetry broken and symmetry protected topological (SPT) phases. These are classified by pairs
$(H,\beta)$,
where, $H \leq G$ is a subgroup representing  the  symmetry 
    \textit{unbroken} in one of the vacua $v$
and $\beta \in H^2(H,U(1))$ is the SPT phase for the unbroken $H$ symmetry in $v$.

This easily follows from the SymTFT $\fZ(\Vec_G)$, which is a 3d Dijkgraaf-Witten (DW) theory with gauge group $G$. The symmetry boundary is chosen to be a Dirichlet b.c.\ (Dir) for the bulk $G$ gauge fields
\begin{equation}\label{DirB}
    \Bsym_{\Vec_{G}} = \cA_\Dir \,.
\end{equation}
Any other topological b.c.\ labeled (Neu$(H),\beta$) is related to this by gauging a subgroup $H \leq G$ with discrete torsion $\beta \in H^2(H,U(1))$. This is realized by imposing Neumann b.c.s for the $H$ gauge fields, which are labeled by  $\beta$. 
By choosing the physical topological boundary to be
\be
\Bphys=\cA_{\Neu(H),\beta}
\ee
we obtain the $G$-symmetric (1+1)d gapped phase associated to the pair $(H,\beta)$, thus reproducing the expected classification. Let us mention two special cases: $H=1$, $\beta=0$ is $\Bphys =\cA_\Dir$ and corresponds to the SSB phase for $G$. The order parameters are untwisted local operators transforming in irreducible representations of $G$, given by all the Wilson lines of the DW theory. On the other hand $H=G$ and any $\beta$, i.e.\ $\Bphys = \cA_{\Neu(G),\beta}$, characterizes the $G$ SPTs. The order parameters are string-like, i.e.\ topological vortices (or magnetic) lines of the DW theory,  dressed by $\beta$-dependent Wilson lines.

\paragraph{$\cS= \Rep(S_3)$.} The simplest example of a group-theoretical non-invertible symmetry is $\Rep(S_3)$, the fusion category of representations of the permutation group $S_3$. We discuss this example in detail in the supplementary materials. Its simple objects are irreducible representations: the trivial 
$\lid$, 1d sign $\lP$ and 2d standard $\lE$ representations, with non-trivial fusions
\be
\lP \otimes \lP = \lid \,,\, \lP \otimes \lE = \lE \,,\, \lE \otimes \lE = \lid \oplus \lP \oplus \lE \,.
\ee
$\Rep(S_3)$ is obtained by gauging the non-anomalous $G=S_3$ symmetry. The SymTFT is the same as that of $\Vec_{S_3}$,
$\fZ(\Rep(S_3))\cong\fZ(\Vec_{S_3})$,
but $\Bsym$ is now Neumann for the bulk gauge fields
\be
\Bsym_{\Rep(S_3)} = \cA_{\Neu} \neq \Bsym_{\Vec_{(S_3)}} = \cA_{\Dir}\,.
\ee
The phases are classified by choosing all possible gapped b.c.s (see (\ref{LagS3})) for $\Bphys$.

\noindent{\bf Trivial Phase:} The SymTFT set-up is 
 \footnote{The topological lines of the SymTFT $\fZ(\Rep(G))$ are labeled by ${\bf Q}_{[c],\rho}$, where $[c]$ is a conjugacy class of $G$ and $\rho$ is an irreducible representation of the centralizer in $G$ of any element $c \in [c]$.}
\begin{equation}\label{eq:s3_dir}
\begin{tikzpicture}
\node at (-2.5,0.5) {$\Bphys= \cA_{\text{Dir}}:$};
\begin{scope}[shift={(0,0)}]
\draw [thick] (0,0) -- (0,1) ;
\draw [thick] (2,0) -- (2,1) ;
\draw [thick] (0, 0.5) -- (2, 0.5) ;
\node[above] at (1,0.5) {$\Q_{[\id],1}$} ;
\node[above] at (0,1) {$\cA_\text{Neu}$}; 
\node[above] at (2,1) {$\cA_\text{Dir}$}; 
\draw [black,fill=black] (0,0.5) ellipse (0.05 and 0.05);
\draw [black,fill=black] (2,0.5) ellipse (0.05 and 0.05);
\end{scope}
\end{tikzpicture}
\end{equation}
We obtain one vacuum, and the full $\Rep(S_3)$ symmetry is (spontaneously) unbroken, thus giving the \textbf{trivial $\Rep(S_3)$-symmetric phase}.

\noindent
{\bf $\Z_2$-SSB Phase:} Neu b.c.\ for $\Z_2\subset \Rep(S_3)$ results in
\be\label{eq:s3_newZ2}
\begin{tikzpicture}
\node at (-2.5,0.25) {$\Bphys= \cA_{\text{Neu}(\bbZ_2)}:$};
\begin{scope}[shift={(0,0)}]
\draw [thick] (0,-0.75) -- (0,1) ;
\draw [thick] (2,-0.75) -- (2,1) ;
\draw [thick] (0,0.5) -- (2,0.5) ;
\draw [thick] (0, -0.25) -- (2, -0.25) ;
\node[above] at (1,0.5) {$\Q_{[\id],1}$} ;
\node[above] at (1,-0.25) {$\Q_{[b],+}$} ;
\node[above] at (0,1) {$\cA_{\text{Neu}}$}; 
\node[above] at (2,1) {$\cA_{\text{Neu}(\bbZ_2)}$}; 
\draw [black,fill=black] (0,0.5) ellipse (0.05 and 0.05);
\draw [black,fill=black] (0,-0.25) ellipse (0.05 and 0.05);
\draw [black,fill=black] (2,0.5) ellipse (0.05 and 0.05);
\draw [black,fill=black] (2,-0.25) ellipse (0.05 and 0.05);
\end{scope}
\end{tikzpicture}
\end{equation}
There are two vacua $v_{1,2}$ exchanged by $\lP$, which is spontaneously broken. Both vacua are on equal footing, with $\lE: v_i \to v_1+v_2 $, and correspondingly no relative Euler term. This is the \textbf{$\bbZ_2$ SSB phase}.

\noindent
{\bf $\Rep(S_3)/\Z_2$ SSB Phase:} Neu b.c.\ for $\Z_3$ results in
\begin{equation}\label{eq:s3_newZ3}
\begin{tikzpicture}
\node at (-2.5,0) {$\Bphys = \cA_{\text{Neu}(\bbZ_3)}:$};
\begin{scope}[shift={(0,0)}]
\draw [thick] (0,-1) -- (0,1) ;
\draw [thick] (2,-1) -- (2,1) ;
\draw [thick] (2,0.5) -- (0,0.25) ;
\draw [thick] (2, 0) -- (0, 0.25) ;
\draw [thick] (0, -0.75) -- (2, -0.75) ;
\node[above] at (1,0.5) {$\Q_{[a],1}$} ;
\node[above ] at (1,-0.75) {$\Q_{[\id],1}$} ;
\node[above] at (0,1) {$\cA_{\text{Neu}}$}; 
\node[above] at (2,1) {$\cA_{\text{Neu}(\bbZ_3)}$}; 
\draw [black,fill=black] (2,0.5) ellipse (0.05 and 0.05);
\draw [black,fill=black] (2,0) ellipse (0.05 and 0.05);
\draw [black,fill=black] (2,-0.75) ellipse (0.05 and 0.05);
\draw [black,fill=black] (0,0.25) ellipse (0.05 and 0.05);
\draw [black,fill=black] (0,-0.75) ellipse (0.05 and 0.05);
\end{scope}
\end{tikzpicture}
\end{equation}
There are two distinct b.c.s for $\Q_{[a], 1}$ in $\cA_{\Neu(\Z_3)}$.
The resulting phase has three vacua, in each of which  $\lP$ is unbroken and $\lE$ is spontenously broken. 
This is the \textbf{$\Rep(S_3) / \bbZ_2$ SSB phase}.
The $\lE$ action takes a vacuum to the sum of the other two vacua. Therefore, all the vacua are again on an equal footing, with no relative Euler terms. This phase and the previous one are examples where spontaneous breaking of non-invertible symmetries \textit{does not} generate physically distinguishable vacua.

\noindent{\bf $\Rep(S_3)$ SSB Phase:} The SymTFT set-up is 
\begin{equation}\label{eq:s3_neu}
\begin{tikzpicture}
\node at (-2.5,0) {$\Bphys = \cA_\text{Neu}:$};
\begin{scope}[shift={(0,0)}]
\draw [thick] (0,-1) -- (0,1) ;
\draw [thick] (2,-1) -- (2,1) ;
\draw [thick] (2,0.7) -- (0,0.7) ;
\draw [thick] (2, 0) -- (0, 0) ;
\draw [thick] (0, -0.7) -- (2, -0.7) ;
\node[above] at (1,0.7) {$\Q_{[a],1}$};
\node[above ] at (1,-0.7) {$\Q_{[\id],1}$};
\node[above] at (1,0) {$\Q_{[b],+}$} ;
\node[above] at (0,1) {$\cA_{\text{Neu}}$}; 
\node[above] at (2,1) {$\cA_{\text{Neu}}$}; 
\draw [black,fill=black] (2,0.7) ellipse (0.05 and 0.05);
\draw [black,fill=black] (2,0) ellipse (0.05 and 0.05);
\draw [black,fill=black] (2,-0.7) ellipse (0.05 and 0.05);
\draw [black,fill=black] (0,0.7) ellipse (0.05 and 0.05);
\draw [black,fill=black] (0,0) ellipse (0.05 and 0.05);
\draw [black,fill=black] (0,-0.7) ellipse (0.05 and 0.05);
\end{scope}
\end{tikzpicture}
\end{equation}   
This phase has three vacua. Two of them, $v_1$ and $v_2$, are permuted among themselves by $\lP$, while one, $v_0$, is left invariant, so under $\Z_2$ the phase decomposes as 
$(\text{Trivial Phase})~\oplus~(\text{$\Z_2$ SSB Phase})$. However, 
$\lE$ permutes these two sub-phases, and hence the three vacua combine into an irreducible phase under $\Rep(S_3)$. This is the $\Rep(S_3)$ \textbf{SSB phase}. Clearly, the three vacua are physically distinct: the action of $\lP$ on $v_0$ is fundamentally different from its action on $v_1,v_2$. Correspondingly, there are relative Euler terms between $v_0$ and $v_1$, and  $v_0$ and $v_2$, but not between $v_1$ and $v_2$. This phase is an example where spontaneous breaking of non-invertible symmetry \textit{does} generate physically distinguishable vacua!

\paragraph{$\cS= \Ising$.} One of the simplest examples of a non-group-theoretical symmetry is the $\Ising$ symmetry,
which is also the Tambara-Yamagami $\TY(\Z_2)$. The simple objects are
$\lid$, $\lP$ (generating a $\Z_2$) and the non-invertible  $\lS$ with  non-trivial fusions
\be
 \lP\otimes \lP = \lid\,,\ \lP \otimes \lS =\lS\,,\quad  \lS \otimes \lS = \lid \oplus \lP\,.
\ee
The SymTFT in this case is the 3d TQFT carrying modular fusion category $\cZ(\Ising)=\Ising\boxtimes\overline\Ising$
of topological line defects. This SymTFT admits only one topological b.c., which we denote by
\be
\Bsym_\Ising=\cA_{\text{Dir},\Neu} \,,
\ee
where the notation follows the one for the subsequent case $\cS = \TY(\bbZ_N)$, of which $\Ising$ is $N=2$.
Hence the only possible configuration is \footnote{The notation for the $\cZ(\Ising)$ lines follows the one in \cite{Bhardwaj:2023ayw}. The relevant aspect to notice here is that there are three bulk topological line defects that can end on both boundaries.}

\begin{equation}
\begin{tikzpicture}
\node at (-2.5,0) {$\Bphys=\cA_{\text{Dir},\Neu}:$};
\begin{scope}[shift={(0,0)}]
\draw [thick] (0,-1) -- (0,1) ;
\draw [thick] (2,-1) -- (2,1) ;
\draw [thick] (2,0.7) -- (0,0.7) ;
\draw [thick] (2, 0) -- (0, 0) ;
\draw [thick] (0, -0.7) -- (2, -0.7) ;
\node[above] at (1,0.7) {$\Q_{\id+\lP}$};
\node[above ] at (1,-0.7) {$\Q_{\id,+}$};
\node[above] at (1,0) {$\Q_{\id,-}$} ;
\node[above] at (0,1) {$\cA_{\text{Dir},\Neu}$}; 
\node[above] at (2,1) {$\cA_{\text{Dir},\Neu}$}; 
\draw [black,fill=black] (2,0.7) ellipse (0.05 and 0.05);
\draw [black,fill=black] (2,0) ellipse (0.05 and 0.05);
\draw [black,fill=black] (2,-0.7) ellipse (0.05 and 0.05);
\draw [black,fill=black] (0,0.7) ellipse (0.05 and 0.05);
\draw [black,fill=black] (0,0) ellipse (0.05 and 0.05);
\draw [black,fill=black] (0,-0.7) ellipse (0.05 and 0.05);
\end{scope}
\end{tikzpicture}
\end{equation}
This phase has three vacua. Similarly to the $\Rep(S_3)$ SSB phase, this $\Ising$-symmetric phase decomposes as $
(\text{Trivial Phase})\oplus(\text{$\Z_2$ SSB Phase})$
under the $\Z_2$, with the two sub-phases permuted by the non-invertible symmetry $\lS$. The three vacua form an irreducible phase under $\Ising$. We call this the \textbf{$\Ising$ SSB phase}.
This is another example where spontaneous breaking of non-invertible symmetry generates physically distinguishable vacua. Despite the similarity with the $\Rep(S_3)$ SSB, however, this phase carries {\it different} relative Euler terms.

\paragraph{$\cS= \TY(\Z_N)$.} The $\TY (\Z_N)$ symmetry involves a $\Z_N$ subsymmetry $\lA^i$, $i=1,\dots,N$ and a non-invertible symmetry $\lS$, with fusion rules 
\begin{equation}\label{eq:TYN_fusion}
    \lA^N \cong \lid \quad,\quad \lA \otimes \lS \cong\lS\ot\lA\cong \lS \quad,\quad \lS \otimes \lS = \oplus_{i=1}^{N} \lA^i \,.
\end{equation}
The SymTFT $\fZ(\TY (\Z_N))$ can be obtained by gauging $\Z_2$ electric-magnetic duality symmetry of 3d $\Z_N$ DW gauge theory \cite{Kaidi:2022cpf}. 
The symmetry boundary is fixed to be 
\be
\Bsym_{\TY(\Z_N)}=\cA_{\Dir,\Neu} \,,
\ee
which can be understood as a decomposition into $\Z_N$ DW theory boundaries
$\cA_{\Dir,\Neu}\equiv\cA_{\Dir}\oplus \cA_{\Neu}$.

\noindent{\bf $\TY(\Z_N)$ SSB = $(\Z_1,\Z_N)$ SSB Phase:} 
Choosing 
\be\begin{tikzpicture}
\node at (-2.5, 0.25) {$\Bphys=\Bsym_{\TY(\Z_N)}=\cA_{\Dir,\Neu}:$} ;
\begin{scope}[shift={(0,0)}]
\draw [thick] (0,-0.75) -- (0,1) ;
\draw [thick] (2,-0.75) -- (2,1) ;
\draw [thick] (0,0.5) -- (2,0.5) ;
\draw [thick] (0, -0.25) -- (2, -0.25) ;
\node[above] at (1,0.5) {$\Q_{0,\pm}$} ;
\node[above] at (1,-0.25) {$\Q_{e,0}$} ;
\node[above] at (0,1) {$\cA_{\Dir,\Neu}$}; 
\node[above] at (2,1) {$\cA_{\Dir,\Neu}$}; 
\draw [black,fill=black] (0,0.5) ellipse (0.05 and 0.05);
\draw [black,fill=black] (0,-0.25) ellipse (0.05 and 0.05);
\draw [black,fill=black] (2,0.5) ellipse (0.05 and 0.05);
\draw [black,fill=black] (2,-0.25) ellipse (0.05 and 0.05);
\end{scope}
\end{tikzpicture}
\ee
with $e = {1,2,\dots, N-1}$, we obtain $N+1$ vacua. This phase decomposes into 
$\text{($\Z_N$-Trivial)}\oplus\text{($\Z_N$-SSB)}$
phase in terms of the $\Z_N$ subsymmetry, with one vacuum participating in the $\Z_N$ trivial phase and the remaining $N$ participating in the $\Z_N$ SSB phase. The two sub-phases are exchanged by the non-invertible $\lS$, with the presence of Euler terms.

\noindent{\bf $(\Z_p,\Z_q)$ SSB Phases:} If $N=pq$ for some $p,q \in \Z_{>1}$, one finds a new algebra $\cA_{\Neu(\Z_p,\Z_q)}$, which can be decomposed using the initial DW theory as
$\cA_{\Neu(\Z_p,\Z_q)}\equiv\cA_{\Neu(\Z_p)}\oplus\cA_{\Neu(\Z_q)}$, $p\le q$.

The SymTFT set-up is 
\begin{equation}
\begin{tikzpicture}
\node at (-2.5,0) {$\Bphys=\cA_{\Neu(\Z_p,\Z_q)}:$};
\begin{scope}[shift={(0,0)}]
\draw [thick] (0,-1.75) -- (0,1) ;
\draw [thick] (2,-1.75) -- (2,1) ;
\draw [thick] (2,0.7) -- (0,0.7) ;
\draw [thick] (2, 0) -- (0, 0) ;
\draw [thick] (0,-1.05) -- (2, -0.7);
\draw [thick] (0,-1.05) -- (2, -1.4) ;
\node[above] at (1,0.7) {$\Q_{0,\pm}$};
\node[above] at (1,0) {$\Q_{e_1,0}$} ;
\node[above ] at (1,-0.9) {$\Q_{e_2,0}$};
\node[above ] at (1,-1.8) {$\Q_{e_2,0}$};
\node[above] at (-0.2,1) {$\cA_{\Dir,\Neu}$}; 
\node[above] at (2.2,1) {$\cA_{\text{Neu}(\bbZ_p, \Z_q)}$}; 
\draw [black,fill=black] (2,0.7) ellipse (0.05 and 0.05);
\draw [black,fill=black] (2,0) ellipse (0.05 and 0.05);
\draw [black,fill=black] (2,-0.7) ellipse (0.05 and 0.05);
\draw [black,fill=black] (2,-1.4) ellipse (0.05 and 0.05);
\draw [black,fill=black] (0,0.7) ellipse (0.05 and 0.05);
\draw [black,fill=black] (0,0) ellipse (0.05 and 0.05);
\draw [black,fill=black] (0,-1.05) ellipse (0.05 and 0.05);
\end{scope}
\end{tikzpicture}    
\end{equation}
where $e_i$ label lines that can end $i$ times.
This phase has $(p+q)$ vacua and decomposes into $\text{($\Z_p$-Trivial)}\oplus\text{($\Z_q$-SSB)}$ under the $\bbZ_N$ generator.

\smallskip

\noindent
{\bf 4d Application.} 
In $d>2$ \footnote{The higher dimensional case is only outlined here, and will be discussed in detail in \cite{BBPSd}.} OPs will be both local and extended operators and there can also be non-trivial topological order. 
SymTFT prescription characterizes e.g.\ the IR gapped phases of 4d QFTs with 0- and 1-form symmetries. A classic instance is  $\mathcal{N}=1$ Super-Yang-Mills  with $SU(2)$ gauge group. This has a $\Z_4^{(0)}$ spontaneously broken to $\Z_2^{(0)}$, resulting in two vacua. The theory has also a  $\bbZ_2^{(1)}$ 1-form symmetry unbroken in both vacua, as signaled by the Wilson lines, its OPs, having area law. 
Gauging the 1-form symmetry gives the $SO(3)_+$ theory \footnote{Without loss of generality we can focus on this global variant, with the treatment for $SO(3)_-$, obtained by stacking an additional 1-form symmetry SPT before its gauging, being completely analogous.}, which also has two vacua, with the difference that $\bbZ_2^{(1)}$ is spontaneously broken in one vacuum and unbroken in the other. Correspondingly, the OPs ('t Hooft lines) have perimeter law in one vacuum and area law in the other. 
The SymTFT is $S = \int_{M_5} \frac{2\pi}{4} a_3 \delta a_1 + \frac{2\pi}{2} c_2 \delta b_2 + \frac{2 \pi}{4} a_1 b_2 b_2$
, where the last term is due to a mixed-anomaly between $\bbZ_4^{(0)}$ and $\bbZ_2^{(1)}$. Notice $\bbZ_2^{(0)} \subset \bbZ_4^{(0)}$ is anomaly free.
The topological defects of the SymTFT are the  Wilson `surfaces' $\Q_x$, with $x$ a  gauge field in $S$ (for details, see \cite{Apruzzi:2022rei,Kaidi:2023maf}).

\noindent 
{\bf SymTFT for the IR of the $SU(2)$:} on $\Bsym_{SU(2)}$ we impose Dirichlet b.c.\ for $a_1$, i.e.\  $\Q_{a_1}$ terminates, while on $\Bphys$ we choose Neu$(\bbZ_2)$, so only $\Q_{a_1}^2$, generating $\bbZ_2^{(0)} \subset \bbZ_4^{(0)}$, terminates.
The only non-trivial line ending on both boundaries is $\Q_{a_1}^2$, so after compactification we obtain one non-trivial untwisted local operator, and hence two vacua $v_\pm$. These are permuted by $\Q_{a_3}$, so we have a $\bbZ_2^{(0)}$ SSB phase. Let us now turn to the 1-form symmetry. We claim $SU(2)$ is realized by choosing for $\Q_{b_2}$ Dirichlet b.c.\  on $\Bsym$ and Neumann b.c.\ on $\Bphys$ (and consequently the opposite for $\Q_{c_2}$). Correspondingly, after compacitifying we obtain no untwisted line operator form the surfaces ending on $\Bphys$. This means that in this gapped phase there are no non-trivial line operators in the IR, and therefore $\bbZ_2^{(1)}$ is left unbroken.

{\bf SymTFT for the IR of the $SO(3)$:} $SO(3)_+$ can be obtained by turning the b.c.\ for $\Q_{b_2}$ and $\Q_{c_2}$ into Neumann and Dirichlet respectively on $\Bsym$. This implies that the 4d gapped phase has one non-trivial untwisted line operator coming from the ends of  $\Q_{c_2}$. The fact that we have a line in the IR means that $\bbZ_2^{(1)}$ is spontaneously broken in one vacuum, say $v_+$. We can transition to the other vacuum $v_-$ by applying $\Q_{a_3}$. This turns $\Q_{c_2}$ into a twisted-sector line, attached to a magnetic surface $\Q_{b_2}$. We then have no non-trivial untwisted line, and we conclude $\bbZ_2^{(1)}$ is unbroken in $v_-$. 

\noindent{\bf Conclusions and Outlook.} We proposed a general framework to extend the Landau paradigm of symmetry breaking to categorical symmetries $\cS$ in any dimension, using as key tool the SymTFT for $\cS$ with two gapped boundaries. The SymTFT topological defects that end on the physical boundary are the generalized charges of $\cS$ and provide  the order parameters of the symmetry breaking. 
We illustrated the power of this approach in (1+1)d and $(1+3)$d, and expect many further higher-dimensional applications. This set-up also provides a starting point to explore lattice models with categorical symmetries, as well as phase transitions between gapped $\cS$-symmetric phases.

%%%%%%%%%%%%%%%%%%%%%%%%%%%%%%%%%%%%
%%%%%%%%%%%%%%%%%%%%%%%%%%%%%%%%%%%%
\newpage
%\small 
%\baselineskip=.94\baselineskip
% \let\bbb\bibitem\def\bibitem{\itemsep4pt\bbb}
\bibliography{GenSym}

%%%%%%%%%%%%%%%%%%%%%%%%%%%%%%%%%%%%
%%%%%%%%%%%%%%%%%%%%%%%%%%%%%%%%%%%%

\onecolumngrid

\vspace{0.2cm}
\noindent
\textbf{Acknowledgements.}
We thank Andrea Antinucci, Federico Bonetti, Mathew Bullimore, Nils Carqueville, Jing-Yuan Chen, Christian Copetti, Michele del Zotto, Giovanni Galati, Jonathan Heckman, Wenjie Ji, Anatoly Konechny, Giovanni Rizi, Apoorv Tiwari, Gerard Watts, Jingxiang Wu and Yunqin Zheng for discussions.  LB thanks the ``Triangle Seminar'' in London, the ``FPUK 2024'' meeting in Durham, and the ``Categorical Aspects of Symmetry Program'' at Nordita (funded by the Simons Collaboration on Global Categorical Symmetries) for stimulating interactions. 
This work is supported in part by the  European Union's Horizon 2020 Framework through the ERC grants 682608 (LB, SSN) and 787185 (LB). LB is funded as a Royal Society University Research Fellow through grant
URF{\textbackslash}R1\textbackslash231467. The work of SSN is supported by the UKRI Frontier Research Grant, underwriting the ERC Advanced Grant "Generalized Symmetries in Quantum Field Theory and Quantum Gravity” and the Simons Foundation Collaboration on ``Special Holonomy in Geometry, Analysis, and Physics", Award ID: 724073, Schafer-Nameki.

%%%%%%%%%%%%

\appendix
\newpage

\section{Supplementary Material}
% No page limit

We provide here some of the technical details that support the claims in the main paper. To give a concrete implementation of our general proposal of characterizing symmetric topological phases, we here go through all the steps outlined in the main text for the case of 
\be
\cS = \Rep (S_3) \,.
\ee

\subsection{SymTFT and Drinfeld Center.}

\subsubsection{General Group-Theoretical Symmetry.}

Group-theoretical non-invertible symmetries are those that can obtained by gauging invertible symmetries. This means that (1+1)d gapped phases for group-theoretical symmetries can all be obtained by gauging gapped phases with group symmetries. In such a case, the non-invertible symmetry $\cS$ shares the same SymTFT with its sibling invertible symmetry $\Vec^\omega_G$ (where $G$ is the symmetry group and $\omega\in H^3(G, U(1))$ is the 't Hooft anomaly)
\be
\fZ(\cS)\cong\fZ(\Vec^\omega_G)  \,,
\ee
and hence the same set of generalized charges is also shared between them. Nevertheless, what differentiates the SymTFT with two different symmetries is its b.c.s: what we label as Dirichlet b.c.\ corresponds to $\Vec^\omega_G$ symmetry, whereas various Neumann b.c.s correspond to some parts of the $\Vec^\omega_G$ symmetry being gauged. Ultimately, having different b.c.s translates to allowing different topological line defects (or generalized charges) to condense on the boundaries, which in turn affects which associated irreducible multiplets of local operators charged under $\cS$ are present. Such a multiplet may contain both twisted and untwisted local operators. 

From this point on, we restrict our attention to $\omega=0$. Analogous statements for $\omega\neq0$ can be found in \cite{Bhardwaj:2023idu}. Categorically speaking, topological line defects of $\fZ\left(\Vec_G\right)$ form a modular fusion category, the Drinfeld center $\cZ(\cS)\cong\cZ(\Vec_G)$. The simple lines (or objects) of $\cZ(\Vec_G)$ are
\be
\Q_{[g],R} \,,
\ee
and can be labeled by a conjugacy class $[g]$ of $G$ and an irreducible representation $\boldsymbol{R}$ of $H_g$, the centralizer of any element $g\in [g]$.

For the trivial conjugacy class $[g]=[\id]$, $\boldsymbol{R}$ are irreducible representations of $G$. Physically, these lines $\Q_{[\id],\boldsymbol{R}}$ are the Wilson lines for the 3d $G$ gauge theory, forming the objects of $\Rep(G)$. On the other hand, a line $\Q_{[g],1}$ is a vortex line around which we have a holonomy for the $G$ gauge fields. The remaining lines $\Q_{[g],\boldsymbol{R}}$ are mixed lines obtained by dressing vortex lines with Wilson lines.

\subsubsection{$\Rep(S_3)$ Symmetry}

The simplest example of group-theoretical non-invertible symmetry is $\Rep(S_3)$, the fusion category formed by finite-dimensional representations of the permutation group $S_3$. 
We can easily construct the SymTFT, as it is the same of $\Vec_{S_3}$. To do so we first label the elements of $S_3$ as
\be
S_3=\left\{ \id, a, a^2, b, ab, a^2 b \right\} \quad \text{with} \quad a^3=b^2=\id  \quad \text{and}\quad a b = b a^2 \,.
\ee
As described above, for a (non-anomalous) group $G$ in 2d, the topological defects of the SymTFT are labeled by two pieces of data: a conjugacy classes $[g]$ of $G$ and irreducible representations (irreps) of the centralizer $H_g$ in $G$ of any element $g \in [g]$. In the case of $G = S_3$, one finds the following conjugacy classes
\begin{equation}
    [\id] = \{\id\} \quad , \quad [a] = \{ a, a^2 \} \quad , \quad [b] = \{ b, ab, a^2 b\} \,,
\end{equation}
with corresponding centralizers
\begin{equation}
    H_{\id} = S_3 \quad , \quad H_{a} = \bbZ_3=\{\id,a,a^2\} \quad , \quad H_{b} = \bbZ_2=\{\id,b\} \,.
\end{equation}
Consequently, the simple topological lines of the SymTFT $\fZ(\Vec_{S_3}) = \fZ(\Rep(S_3))$ are labeled by the conjugacy class and centralizer irreps, 
\be
  \Q_{[\id],\boldsymbol{R}} : \quad \boldsymbol{R} = 1,P,E \,, \qquad
  \Q_{[a],\boldsymbol{R}} : \quad \boldsymbol{R} = 1,\omega,\omega^2 \,, \qquad
    \Q_{[b],\boldsymbol{R}} : \quad \boldsymbol{R} = +,- \,,
\ee
where $P$ is the sign irrep of $S_3$, $E$ the 2d irrep of $S_3$, $\omega = e^{\pm 2 \pi i/3}$ is a $\Z_3$ irrep, and $+$, $-$ denote the $\Z_2$ irreps.

\subsection{Lagrangian Algebras.}
Irreducible topological b.c.s of $\fZ(\cS)$ are described by Lagrangian algebras of the category $\cZ(\cS)$. A discussions of Lagrangians can be found in \cite{Kong:2009inh, davydov2013witt, Kong:2013aya, davydov2021braided, Johnson-Freyd:2021chu, Zhao:2022yaw}.
A Lagrangian algebra $\cA$ can be expressed as
\be
\cA=\bigoplus_A n_A \Q_{A} \,, \qquad n_A \in \mathbb{Z}_{\geq 0}\,,
\ee
where $\Q_A$ are simple objects of $\cZ(\cS)$. Any such algebra has to satisfy the fusion coefficient and dimension constraints,
\be
n_A n_B \leq \sum_C N_{AB}^C n_C  \quad \text{and} \quad\dim(\cA):=\sum_{A\in \cA } n_A \dim(\Q_A)=\dim^2(\cS)\,,
\ee
where $N_{AB}^C$ are the fusion coefficients sending $A \otimes B \rightarrow C$ in $\cS$ and $\dim(\cS)$ is the quantum dimension of the category, which can be found as
\be
\dim(\cS)=\sqrt{\sum_{a} \dim^2(a)} \,,
\ee
where the sum is over all simple objects $a$ of $\cS$ and $\dim(a)$ denotes the expectation value of a loop $a$. Finally, we note that all $\Q_A$ participating in $\cA$ must be bosons. 

The topological lines in $\fZ(\Vec_{S_3}) = \fZ(\Rep(S_3))$ that are bosons are
\be
\Q_{[\id],1},\qquad\Q_{[\id],P},\qquad\Q_{[\id],E},\qquad \Q_{[a],1},\qquad \Q_{[b],+} \,,
\ee
and the Lagrangian algebras they form that satisfy the constraints are
\begin{equation}\label{LagS3}
\begin{aligned}
    \cA_{\text{Dir}} &= \Q_{[\id],1} \oplus \Q_{[\id],P} \oplus 2\Q_{[\id],E} \\
    \cA_{\text{Neu}} &= \Q_{[\id],1} \oplus \Q_{[a],1} \oplus \Q_{[b],+} \\
    \cA_{\text{Neu}(\Z_2)} &= \Q_{[\id],1} \oplus \Q_{[\id],E} \oplus \Q_{[b],+} \\
    \cA_{\text{Neu}(\Z_3)} &= \Q_{[\id],1} \oplus \Q_{[\id],P} \oplus 2\Q_{[a],1} \,.
\end{aligned}
\end{equation}
As briefly mentioned previously, various Neumann b.c.s correspond to taking the Dirichlet b.c., $\cA_{\text{Dir}}$, and gauging different subgroups of the full group symmetry. From the SymTFT perspective, gauging simply means imposing free/Neumann b.c.s on the subset of gauge fields one wishes to gauge. Here, fully gauging the $S_3$ symmetry corresponds to $\cA_{\text{Neu}}$, whereas partially gauging $\Z_2\subseteq S_3$, $\Z_3\subseteq S_3$ results in $\cA_{\text{Neu}(\Z_2)}$, $\cA_{\text{Neu}(\Z_3)}$ respectively.

\subsection{$\Rep(S_3)$ Phases}

We are now finally ready to discuss all the gapped phases with $\Rep(S_3)$ symmetry, a symmetry consisting of two non-trivial simple objects $\lP$ and $\lE$, with fusion rules
\be
\lP \ot \lP = 1\,, \qquad \lP \ot \lE = \lE\,, \qquad \lE \ot \lE = 1 \oplus \lP \oplus \lE\,,
\ee
where $\lP$ generates $\Z_2$ subsymmetry and $\lE$ is a non-invertible line defect.

To implement the $\Rep(S_3)$ symmetry for the SymTFT, we take the symmetry boundary to be
\be
\Bsym_{\Rep(S_3)}=\cA_{\text{Neu}} \,.
\ee
Choosing various physical boundaries in \eqref{LagS3} gives rise to the various $\Rep(S_3)$-symmetric phases. Given 4 choices of physical boundaries in the present case, one finds 4 gapped phases, which we study in detail below.

\subsubsection{Trivial Phase}\label{R3T}
By choosing $\Bphys= \cA_{\text{Dir}}$, one finds that only the bulk line $\Q_{[\id],1}$ may end on both sides of the SymTFT (see \eqref{eq:s3_dir}). Hence one finds a phase with a single, trivial (untwisted) local operator which is the \textbf{trivial phase} with one vacuum, where $\Rep(S_3)$ symmetry is spontaneously preserved.  

This phase is characterized by the coexistence of two different kinds of order parameters. One type of order parameters are multiplets of local operators carrying generalized charge $\Q_{[\id],P}$. Such a multiplet comprises of a local operator in twisted sector for $\lP$ symmetry, and hence such an order parameter is a string order parameter. The other type of order parameters are also of string type and carry  generalized charge $\Q_{[\id],E}$, which is a multiplet comprising only of twisted sector operators, one in the $\lE$-twisted sector and one changing $\lE$ to $\lP$.  All the operators discussed above are uncharged under $\lP$. See \cite{Bhardwaj:2023idu} for more details.

\subsubsection{$\Z_2$ SSB Phase}\label{Z2R3SSB}
By choosing $\Bphys= \cA_{\text{Neu}(\bbZ_2)}$, we end up with two untwisted sector local operators, which translates into finding two vacua in this phase (see \eqref{eq:s3_newZ2}). In comparison to the trivial phase, there is now another, non-trivial, untwisted sector local operator associated to $\Q_{[b],+}$ that we label as $\cO^b$. Additionally, there are two twisted sector local operators $\cO^b_\pm$ that belong to the $\Q_{[b],+}$ multiplet that do not participate in the determination of vacua but become important later as order parameters of the resulting phase.

In general, the $n$ vacua $v_i$ (with $i=1,2,\dots,n$) of the TQFT satisfy the relation
\be
v_iv_j=\delta_{ij}v_i \,,
\ee
where $\delta_{ij}$ is the Kronecker delta. To find such vacua given the untwisted local operators in the theory, one must first obtain their operator algebra, in this case of $\{1,\cO^b\}$. The only non-trivial algebra rule we must specify will be of the form
\be
\cO^b\cO^b=\alpha+\beta\cO^b,\qquad(\alpha,\beta)\in\bC^2-\{(0,0)\} \,.
\ee
To constrain the algebra we can study the action of $\lP$ (and $\lE$) on $\cO^b$,
\be\label{eq:bmult_1minus}
\begin{tikzpicture}
\begin{scope}[shift={(-7,0)}]
\draw [thick,fill] (0.5,0) ellipse (0.05 and 0.05);
\node at (0.5,-0.4) {$\cO^b$};
\draw [thick](1.25,0.5) -- (1.25,-0.5);
\node at (1.5,-0.5) {$\lP$};
\node at (2.25,0) {=};
\begin{scope}[shift={(4,0)}]
\node at (-0.1,-0.5) {$\lP$};
\draw [thick,fill] (0.5,0) ellipse (0.05 and 0.05);
\node at (0.5,-0.4) {$\cO^b$};
\node at (1,-0.25) {,};
\draw [thick](-0.35,0.5) -- (-0.35,-0.5);
\end{scope}
\node at (3,0) {$-$};
\end{scope}
\begin{scope}[shift={(8,2.5)}]
\draw [thick,fill] (-7.5,-2.5) ellipse (0.05 and 0.05);
\node at (-7.5,-2.9) {$\cO^b$};
\draw [thick](-6.75,-2) -- (-6.75,-3);
\node at (-6.5,-3) {$\lE$};
\node at (-5.75,-2.5) {=};
\begin{scope}[shift={(-4.25,-2.5)}]
\draw [thick](-0.5,0) -- (0.5,0);
\node at (0,0.25) {$\lE$};
\draw [thick,fill] (0.5,0) ellipse (0.05 and 0.05);
\node at (0.5,-0.4) {$\cO^b_+$};
\draw [thick](-0.5,0.5) -- (-0.5,-0.5);
\node at (-0.25,-0.5) {$\lE$};
\end{scope}
\end{scope}
\end{tikzpicture}
\ee
which also confirms that $\cO^b$ is charged non-trivially under the $\Z_2$ subsymmetry of $\Rep(S_3)$ generated by $\lP$. 
Consequently, by symmetry, $\beta$ must vanish and by rescaling $\cO^b$ one finds the algebra to be 
\be
\cO^b\cO^b=1 \,.
\ee
The two vacua are then idempotent combinations of the identity local operator and $\cO^b$,
\be
v_0=\frac{1+\cO^b}2\,, \qquad
v_1=\frac{1-\cO^b}2\,.
\ee

Now in order to identify the linking actions of $\lP$ and $\lE$ on the vacua, we first have to identify the linking actions on the local operator $\cO^b$ which are
\be\label{eq:bmult_link}
\begin{tikzpicture}
\begin{scope}[shift={(-4.5,0)}]
\draw [thick,fill] (0,0) ellipse (0.05 and 0.05);
\node at (0,-0.4) {$\cO^b$};
\node at (-1,0) {$\lP$};
\draw [thick] (0,0) ellipse (0.75 and 0.75);
\node at (1.5,0) {=};
\begin{scope}[shift={(3,0)}]
\node at (-0.75,0) {$-$};
\draw [thick,fill] (1.5,0) ellipse (0.05 and 0.05);
\node at (1.5,-0.4) {$\cO^b$};
\node at (-0.3,0) {$\lP$};
\draw [thick] (0.5,0) ellipse (0.5 and 0.5);
\end{scope}
\begin{scope}[shift={(6,0)}]
\node at (-0.75,0) {=};
\node at (0,0) {$-$};
\draw [thick,fill] (0.5,0) ellipse (0.05 and 0.05);
\node at (0.5,-0.5) {$\cO^b$};
\node at (1,0) {,};
\end{scope}
\end{scope}
\begin{scope}[shift={(4.5,0)}]
\draw [thick,fill] (0,0) ellipse (0.05 and 0.05);
\node at (0,-0.4) {$\cO^b$};
\node at (-1,0) {$\lE$};
\node at (1.5,0) {=};
\begin{scope}[shift={(-1.75,0)}]
\begin{scope}[shift={(8,0)}]
\draw [thick](-3,0) -- (-2,0);
\node at (-2.5,0.25) {$\lE$};
\draw [thick,fill] (-2,0) ellipse (0.05 and 0.05);
\node at (-2,-0.5) {$\cO^b_+$};
\end{scope}
\draw [thick] (4.5,0) ellipse (0.5 and 0.5);
\node at (3.75,0) {$\lE$};
\end{scope}
\draw [thick] (0,0) ellipse (0.75 and 0.75);
\node at (5,0) {=};
\node at (5.75,0) {0};
\end{scope}
\end{tikzpicture}
\ee
These linking actions then translate to the following linking actions on the vacua
\be
\ba
\lP\, \bigcirc &: \qquad v_0\to v_1\,,\qquad
v_1\to v_0 \,,\\
\lE \, \bigcirc &: \qquad v_0, v_1\to 1=v_0+v_1 \,,
\ea
\ee
which shows that there are no relative Euler terms between the two vacua as the linking actions only contain trivial factors for all terms. This phase is the \textbf{$\Z_2$ SSB phase} as the $\Z_2$ subgroup symmetry of the overall $\Rep(S_3)$ symmetry is spontaneously broken in both vacua. Note that $\lE$ acts on a vacuum to generate both vacua and hence is also spontaneously broken. However, the two vacua are physically indistinguishable as far as the action of $\Rep(S_3)$ is concerned.

This phase is characterized by the coexistence of two different kinds of order parameters. One is of mixed-type, has generalized charge $\Q_{[b],+}$, and is a 3-dimensional multiplet of local operators: one is untwisted, one is in the twisted sector for $\lE$, and one transitions between lines $\lE$ and $\lP$. All of them are charged non-trivially under $\lP$. The other is of string-type, has generalized charge $\Q_{[\id],E}$, and is a 2-dimensional multiplet of local operators: one is in the twisted sector for $\lE$, while the other transitions between lines $\lE$ and $\lP$. Both are uncharged under $\lP$.

\subsubsection{$\Rep(S_3)/\Z_2$ SSB Phase}\label{Z3R3SSB}
By choosing $\Bphys = \cA_{\text{Neu}(\bbZ_3)}$, interestingly, one finds that $\Q_{[a],1}$ on its own gives rise to 2 untwisted sector local operators, $\cO^a_{+,1}$ and $\cO^a_{+,2}$, after collapsing the SymTFT sandwich (see \eqref{eq:s3_newZ3}). This happens as the bulk line may end twice on the physical boundary in this case which can be seen from its Lagrangian algebra. Including the trivial local operator, this phase includes 3 untwisted local operators and thus 3 vacua. Additionally, there are two twisted local operators $\cO^a_{-,1}$  and $\cO^a_{-,2}$ descending from $\cO^a_{-}$ of the $\Q_{[a],1}$ multiplet, which become additional order parameters of the phase.

In order to find the operator algebra in this setting, we turn our attention to the $\Z_3 \subseteq S_3$ subgroup symmetry localized on the physical boundary $\Bphys$. The $\Z_3$ action has the following transformation property 
\be
\cO^a_{+,1}\to\omega\,\cO^a_{+,1}\,, \qquad
\cO^a_{+,2}\to\omega^2\,\cO^a_{+,2} \,,
\ee
forcing the algebra to take the form
\be
\cO^a_{+,1}\cO^a_{+,1}=\cO^a_{+,2}\,, \qquad
\cO^a_{+,2}\cO^a_{+,2}=\cO^a_{+,1}\,, \qquad
\cO^a_{+,1}\cO^a_{+,2}=1 \,,
\ee
after imposing associativity and rescaling $\cO^a_{+,1}$ and $\cO^a_{+,2}$. This determines the three vacua to be
\be
v_i=\frac{1+\omega^i\,\cO^a_{+,1}+\omega^{2i}\,\cO^a_{+,2}}{3},\qquad i\in\{0,1,2\}\,, \qquad \omega = e^{\pm 2\pi i /3}\,.
\ee

The linking action of $\lP$ on $\cO^a_{+,1}$ and $\cO^a_{+,2}$ is trivial, 
\be\label{paction1}
\lP \, \bigcirc : 
\qquad \cO^a_{+,1}\to \,\cO^a_{+,1}\,, \qquad
\cO^a_{+,2}\to \,\cO^a_{+,2} \,.
\ee
From this one learns that the symmetry $\lP$ leaves each vacuum invariant. Since the $\Z_2$ subsymmetry of $\Rep(S_3)$ is spontaneously unbroken in all three vacua, we refer to this phase as the \textbf{$\Rep(S_3)/\Z_2$ SSB phase}.

On the other hand, the linking action of $\lE$ with the vacua is more interesting as $\lE$ links with $\cO^a_+$ in the following way
\be\label{eq:amult_E_link}
\begin{tikzpicture}
\begin{scope}[shift={(-7,0)}]
\draw [thick] (0,0) ellipse (0.75 and 0.75);
\draw [thick,fill] (0,0) ellipse (0.05 and 0.05);
\node at (0,-0.4) {$\cO^a_+$};
\node at (-1.05,0) {$\lE$};
\node at (1.25,0) {=};
\begin{scope}[shift={(4,0)}]
\node at (-2.35,0) {$-$};
\node at (-1.85,0) {$\half$};
\draw [thick,fill] (0.5,0) ellipse (0.05 and 0.05);
\node at (0.5,-0.5) {$\cO^a_+$};
\draw [thick] (-0.5,0) ellipse (0.5 and 0.5);
\node at (-1.35,0) {$\lE$};
\node at (1.25,0) {+};
\begin{scope}[shift={(7.5,0)}]
\node at (-5.15,0) {$\left(\omega+\half\right)$};
\node at (-4.05,0) {$\lE$};
\draw [thick](-2.75,0) -- (-1.75,0);
\node at (-2.25,0.25) {$\lP$};
\draw [thick,fill] (-1.75,0) ellipse (0.05 and 0.05);
\node at (-1.75,-0.5) {$\cO^a_-$};
\draw [thick] (-3.25,0) ellipse (0.5 and 0.5);
\end{scope}
\end{scope}
\end{scope}
\begin{scope}[shift={(6.25,-1)}]
\node at (-2.75,1) {=};
\node at (-2.25,1) {$-$};
\node at (-1.75,1) {$\frac{2}{2}$};
\draw [thick,fill] (-1,1) ellipse (0.05 and 0.05);
\node at (-1,0.5) {$\cO^a_+$};
\node at (-0.375,1) {$+$};
\node at (0.25,1) {$0$};
\end{scope}
\begin{scope}[shift={(10,-1)}]
\node at (-2.85,1) {=};
\node at (-2.15,1) {$-$};
\draw [thick,fill] (-1.5,1) ellipse (0.05 and 0.05);
\node at (-1.5,0.5) {$\cO^a_+$};
\end{scope}
\end{tikzpicture}
\ee
where the second term on the right-hand side vanishes because there are no topological local operators in $\Rep(S_3)$ converting the line $\lP$ into the identity line.

Thus the only non-trivial linking is
\be
\lE \, \bigcirc : 
\qquad \cO^a_{+,1}\to-\,\cO^a_{+,1}\,, \qquad
\cO^a_{+,2}\to-\,\cO^a_{+,2} \,,
\ee
which implies its linking action on the vacua is
\be
\lE \, \bigcirc:  \quad v_0\to v_1+v_2\,, \qquad
v_1\to v_2+v_0\,, \qquad
v_2\to v_0+v_1 \,,
\ee
hence there are no relative Euler terms between the three vacua. Note that $\lE$ acts on a vacuum to generate the other two vacua, and hence is spontaneously broken. However, all three vacua are physically indistinguishable as far as the action of $\Rep(S_3)$ is concerned.

This phase is again characterized by the coexistence of two types of order parameters. One is a string order parameter discussed above, which carries generalized charge $\Q_{[\id],P}$. The other is of mixed-type, has generalized charge $\Q_{[a],1}$, and is a 2-dimensional multiplet comprising of an untwisted sector local operator and an $\lP$-twisted sector local operator. The two operators are mixed into each other by the action of $\lE$, but are uncharged under $\lP$.

\subsubsection{$\Rep(S_3)$ SSB Phase}\label{R3SSB}
Choosing $\Bphys = \cA_\text{Neu}$ results in a phase with 3 untwisted local operators and thus 3 vacua (see \eqref{eq:s3_neu}). We will call these 3 untwisted local operators 1,  $\cO^a_+$ and $\cO^b$, which descent from $\Q_{[\id],1}$, $\Q_{[a],1}$, and $\Q_{[b],+}$, respectively. Additionally, there are again some twisted local operators that we will mention later when we discuss order parameters.

To determine the operator algebra in this case, we first look at the bulk fusion rule
\be
\Q_{[b],+}\ot \Q_{[a],1} \cong \Q_{[b],+}\oplus\Q_{[b],-}
\ee
which after noticing that $\Q_{[\lid],1}$ and $\Q_{[a],1}$ are absent on the RHS means that after a possible rescaling of $\cO^b$ one must find
\be\label{pr1}
\cO^a_+\cO^b=\cO^b\,.
\ee

Similarly, since the fusion
\be
\Q_{[a],1}\ot \Q_{[a],1} \cong \Q_{[\id],1}\oplus\Q_{[\id],P} \oplus \Q_{[a],1}
\ee
does not contain $\Q_{[b],+}$, we must have
\be\label{pr2}
\cO^a_+\cO^a_+=\alpha+(1-\alpha)\cO^a_+,\qquad\alpha\in\bC\,,
\ee
where the relative weight between the two coefficients on the RHS has been set by imposing associativity with (\ref{pr1}).

To further constrain $\alpha$, we first establish the action of $\lE$ on $\cO^a_+$,
\be\label{act1}
\begin{tikzpicture}
\begin{scope}[shift={(1.25,0)}]
\draw [thick](-0.5,0) -- (0.5,0);
\node at (0,0.25) {$\lP$};
\draw [thick,fill] (0.5,0) ellipse (0.05 and 0.05);
\node at (0.6,-0.4) {$\cO^a_-$};
\draw [thick](-0.5,0.5) -- (-0.5,-0.5);
\node at (-0.25,-0.5) {$\lE$};
\end{scope}
\begin{scope}[shift={(-7.5,0)}]
\draw [thick,fill] (0.5,0) ellipse (0.05 and 0.05);
\node at (0.5,-0.4) {$\cO^a_+$};
\draw [thick](1.25,0.5) -- (1.25,-0.5);
\node at (1.5,-0.5) {$\lE$};
\node at (2.25,0) {=};
\begin{scope}[shift={(4.5,0)}]
\node at (-0.25,-0.5) {$\lE$};
\draw [thick,fill] (0.5,0) ellipse (0.05 and 0.05);
\node at (0.5,-0.5) {$\cO^a_+$};
\draw [thick](-0.5,0.5) -- (-0.5,-0.5);
\end{scope}
\end{scope}
\node at (-4.5,0) {$-$};
\node at (-4,0) {$\half$};
\node at (-1.75,0) {+};
\node at (-0.5,0) {$\left(\omega+\half\right)$};
\end{tikzpicture}
\ee
We can then apply the action of $\lE$ on \eqref{pr2} and by matching the $\cO^a_-$ contributions on both sides it can be shown that
\be
\cO^a_-\cO^a_+=-(1-\alpha)\cO^a_- \,.
\ee
Imposing associativity with (\ref{pr2}) fixes $\alpha = \frac{1}{2}$ (as the other root produces inconsistent results).

For the final product relation, note that the fusion
\be
\Q_{[b],+}\ot\Q_{[b],+} \cong \Q_{[\id],1}\oplus\Q_{[\id],E} \oplus \bigoplus_{i=0}^2\Q_{[a],\omega^i}
\ee
does not contain $\Q_{[b],+}$, and so imposing associativity and rescaling we obtain
\be
\cO^b\cO^b=\half+\cO^a_+ \,.
\ee

Putting everything together, the operator algebra consists of the following three non-trivial rules
\be
\cO^a_+\cO^b=\cO^b\,, \qquad
\cO^a_+\cO^a_+=\half(1+\cO^a_+)\,, \qquad
\cO^b\cO^b=\half+\cO^a_+ \,,
\ee
from which one can determine the vacua to be
\be
v_0=\frac23\left(1-\cO^a_+\right)\,, \qquad
v_1=\frac16\left(1+2\cO^a_++\sqrt{6}\cO^b\right)\,, \qquad
v_2=\frac16\left(1+2\cO^a_+-\sqrt{6}\cO^b\right) \,.
\ee

Similarly to \eqref{paction1} and \eqref{eq:amult_E_link}, we deduce again that linking action of $\lP$ is
\be
\lP \, \bigcirc : \qquad \cO^a_+\to\cO^a_+\,, \qquad
\cO^b\to-\cO^b
\ee
and hence it acts on the vacua as
\be\label{linking5}
\lP \, \bigcirc : \qquad v_0\to v_0\,, \qquad
v_1\to v_2\,, \qquad
v_2\to v_1 \,.
\ee
Thus the present phase, which we refer to as the \textbf{$\Rep(S_3)$ SSB phase}, decomposes as a sum of a $\Z_2$ SSB phase (formed by vacua $v_1$ and $v_2$) and a $\Z_2$ non-SSB phase (formed by vacuum $v_0$).

The linking action of $\lE$ on the operators is
\be
\lE \, \bigcirc:  \qquad
\cO^a_+\to-\cO^a_+\,, \qquad
\cO^b\to0
\ee
while the action on the vacua is
\be\label{linking6}
\lE \, \bigcirc:  \quad v_0\to\frac23\left(2+\cO^a_+\right)=v_0+2(v_1+v_2)\,, \qquad
v_1\to\frac13\left(1-\cO^a_+\right)=\half v_0\,, \qquad
v_2\to\frac13\left(1-\cO^a_+\right)=\half v_0 \,.
\ee
Thus, $\lE$ is spontaneously broken.

Judging from \eqref{linking5} and \eqref{linking6}, one can clearly see there are no relative Euler terms between vacua $v_1$ and $v_2$, however,  the presence of fractions in \eqref{linking6} uncovers relative Euler terms between vacua $v_0$ and $v_1$, $v_2$. The vacuum $v_0$ is thus physically distinguishable from the vacua $v_1$ and $v_2$, as is also apparent from the action of the unique $\Z_2$ subsymmetry $\lP$ of $\Rep(S_3)$. In this case, the spontaneous breaking of non-invertible symmetry is linked to the appearance of physically indistinguishable vacua!

This phase is again characterized by the coexistence of two types of order parameters. Both are of mixed type and have been discussed above: one of them carries generalized charge $\Q_{[a],1}$, while the other carries generalized charge $\Q_{[b],+}$.

%%%%%%%%%%%%%%%%%%%%%%%%%%%%%%%%%%%%
%%%%%%%%%%%%%%%%%%%%%%%%%%%%%%%%%%%%

\end{document}